\documentclass[12pt]{article}

\usepackage{scicite}

\usepackage{times}

\usepackage{graphicx}
\usepackage{epstopdf}

\newenvironment{sciabstract}{%
\begin{quote} \bf}
{\end{quote}}

\newcounter{lastnote}

\title{Numerical Simulations Unravel the Cosmic Web}

\author
{Claude-Andr\'e Faucher-Gigu\`ere,$^{1\ast}$ Adam Lidz,$^{1}$ Lars Hernquist$^{1}$\\
\\
\normalsize{$^{1}$Department of Astronomy, Harvard University}\\
\normalsize{60 Garden Street, Cambridge, MA 02138, USA}\\
\\
\normalsize{$^\ast$To whom correspondence should be addressed; E-mail:  cgiguere@cfa.harvard.edu.}
}

\date{}

\newcommand{\Lya}{Ly$\alpha$}

\begin{document} 


\baselineskip24pt


\maketitle


\begin{sciabstract}
The universe is permeated by a network of filaments, sheets, and knots collectively forming a ``cosmic web.''
The discovery of the cosmic web, especially through its signature of absorption of light from distant sources by neutral hydrogen in the intergalactic medium, exemplifies the interplay between theory and experiment that drives science, and is one of the great examples in which numerical simulations have played a key and
decisive role.
We recount the milestones in our understanding of cosmic structure, summarize its impact on astronomy, cosmology, and physics, and look ahead by outlining the challenges faced as we prepare to probe the cosmic web at new wavelengths.
\end{sciabstract}

Cosmologists envision a universe made up of filaments, knots, and
sheets reminiscent of pancakes \cite{1970A&A.....5...84Z}, dubbed
the ``cosmic web'' by Bond and collaborators \cite{1996Natur.380..603B}.
This picture of the cosmos -- now well entrenched even in popular culture -- is, however, much more than mere fantasy.
Indeed, it is one of the best-established results of
cosmological research and underpins much of our contemporary
understanding of large-scale structure and galaxy formation. 

 
After Schmidt's spectroscopic observations of a distant quasar (a fantastically bright point source located at a cosmological distance, later understood to be powered by a supermassive black hole accreting from its host galaxy)
in 1965 \cite{1965ApJ...141.1295S}, Gunn \& Peterson \cite{1965ApJ...142.1633G} and (independently)
Scheuer \cite{1965Natur.207..963S} and Shklovski
\cite{1965SvA.....8..638S} pointed out that the spectra of
distant quasars should show absorption by neutral hydrogen along the
line of sight blueward of their Lyman-$\alpha$ emission line.
Refining this
calculation, Bahcall \& Salpeter \cite{1965ApJ...142.1677B} argued
that the clumpy intergalactic gas should give rise to a collection of
discrete absorption lines, termed the ``Lyman-$\alpha$~(\Lya)~forest.''
The following year, observers, particularly Lynds and coworkers
\cite{1966ApJ...144..447B},
indeed saw those lines.

Their exact nature, however, gave rise to a more extended and heated debate. Were the lines -- then poorly resolved - really of cosmological origin, or were they instead associated with the quasars themselves?  The extreme conditions needed for quasar ejecta to produce the \Lya~forest \cite{1976ComAp...6..133G}, the random redshift distribution of the absorption lines \cite{1969ApJ...156L...7B}, as well as the association of metal-rich systems with intervening galaxies \cite{1986A&A...155L...8B}, eventually left no doubt that most of the \Lya~clouds are in fact cosmological.

Inspired by research on interstellar medium, the \Lya~absorbers were
originally envisioned as dense cool clouds confined by pressure within
a hotter, tenuous intercloud background \cite{1980ApJS...42...41S,
1983ApJ...268L..63O}. But this model raised questions of its own,
namely how did the clouds form in the first place? Ultimately, the
model simply failed to account for the observed distribution of
neutral hydrogen column densities, and other confinement mechanisms
involving gravity 
were also found unsatisfactory.

\begin{figure}[ht]
\begin{center}
\includegraphics[width=0.95\textwidth]{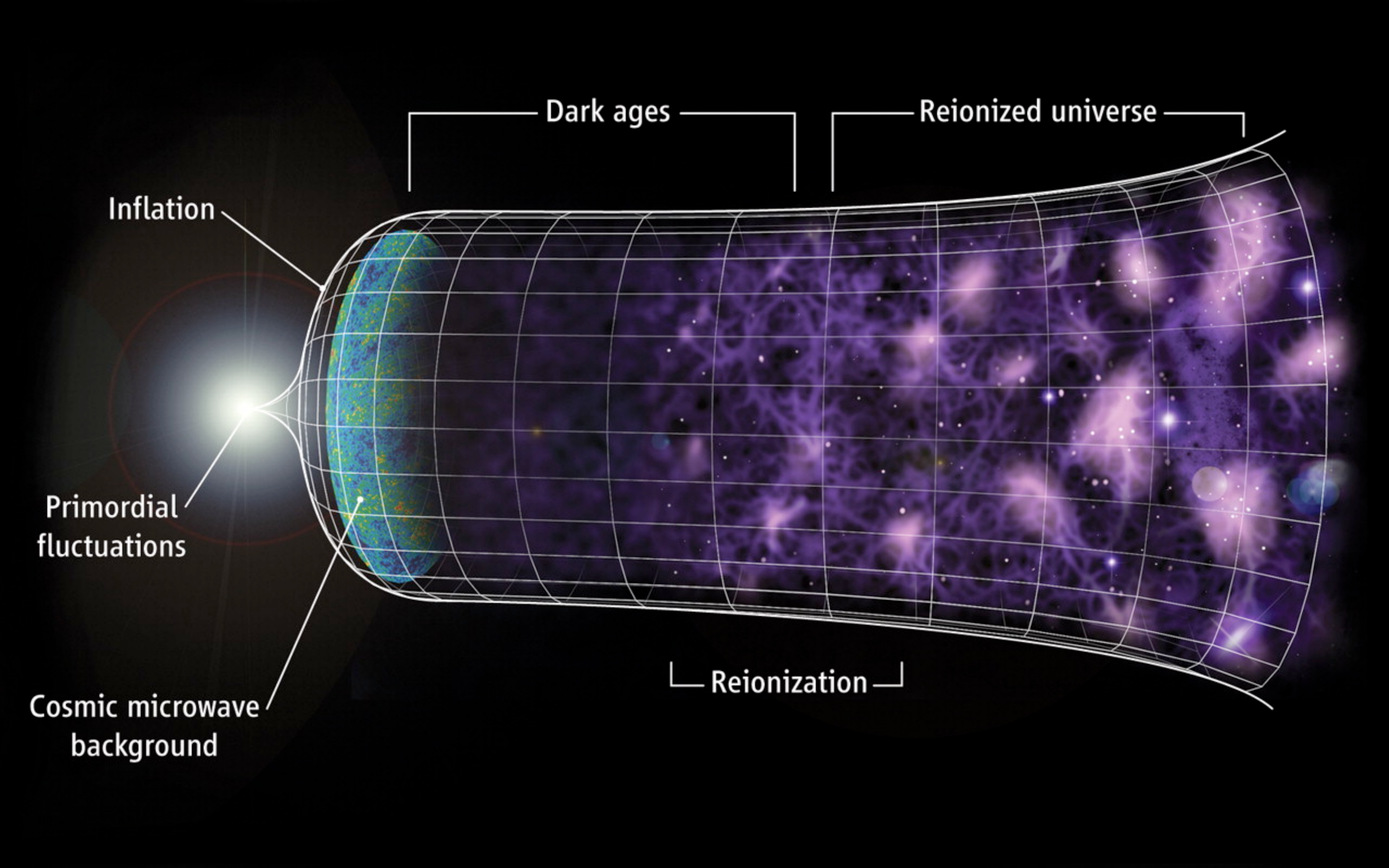}
\end{center}
\caption{View of the expanding universe illustrating the evolving cosmic web. Primordial fluctuations of quantum mechanical origin are stretched by an early superluminal phase of expansion known as "inflation." Four hundred thousand years after the Big Bang, these fluctuations imprint tiny anisotropies in the cosmic microwave background as electrons and protons combine to form neutral hydrogen. As the universe continues to expand, these fluctuations grow through the gravitational instability, eventually giving rise to the first stars and galaxies, whose radiation reionizes the intergalactic medium, thereby ending the cosmological "dark ages." The cosmic web sharpens as the universe becomes more mature and becomes visible in the Ly forest and in the spatial distribution of galaxies.}
\label{expanding universe}
\end{figure}

The key breakthrough in elucidating the nature of the \Lya~forest came
as astrophysicists tried to understand the formation of structure in the universe. In our current model of structure formation, tiny fluctuations
in the primordial plasma grew through the gravitational instability,
eventually forming a cosmic network of knots, filaments, and sheets.
As gravity is a purely attractive force, regions of slightly higher density in the
early universe accrete matter from their surroundings and grow more
overdense with time. 
As the universe evolves, the cosmic web sharpens: underdense regions, known as voids, empty material onto the filaments, and this material subsequently flows into overdense knots (Fig. \ref{expanding universe}). On large scales, where gas pressure can be neglected, intergalactic neutral hydrogen should trace this web.

What if, then, the \Lya~forest was simply a representation of the cosmic density field? 
This was a far-reaching question transcending intergalactic clouds,
for its answer would not only provide a powerful test of structure
formation models, but would also potentially leave us with a
new and exquisitely detailed astrophysical and cosmological probe.

\begin{figure}[ht]
\begin{center}
\includegraphics[width=0.65\textwidth]{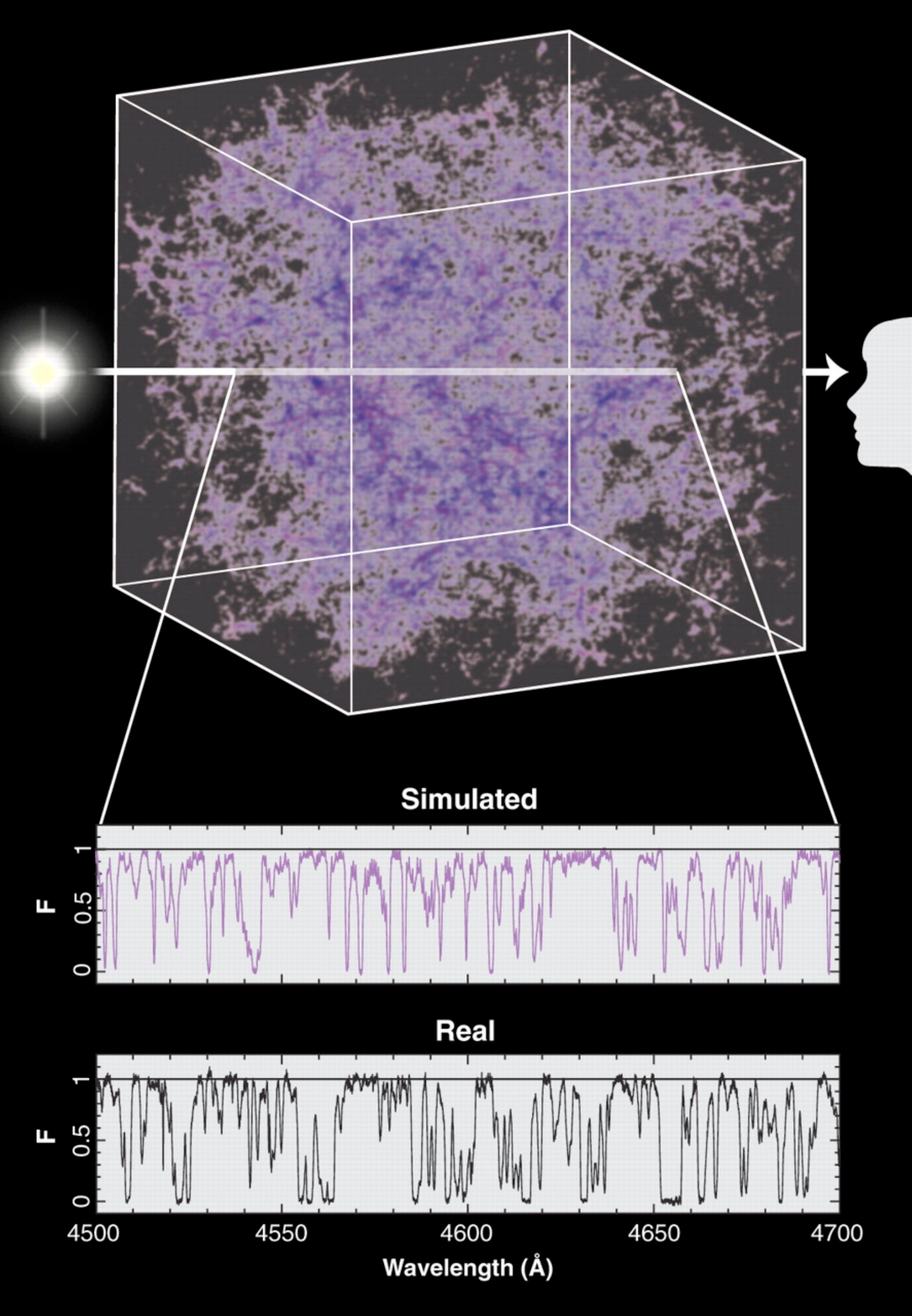}
\end{center}
\caption{Illustration of the \Lya~forest.
An observer looks at a distant quasar.
Neutral hydrogen tracing the cosmic web produces absorption features, collectively known as the \Lya~forest, in the quasar spectrum.
The figure shows a line of sight through a cosmological simulation, with the resulting mock \Lya~forest compared to the spectrum of an actual quasar, known as Q1422 and located at redshift $z=3.6$ (spectrum courtesy of M.~Rauch, Observatories of the Carnegie Institution of Washington, Pasadena, CA, and W.~Sargent, California Institute of Technology, Pasadena, CA).
The similarity between the mock and actual spectra is remarkable, unambiguously elucidating the nature of the \Lya~forest as the imprint of cosmological fluctuations.
Each spectrum has been normalized by its continuum level in this figure.}
\label{lya forest}
\end{figure}

After promising analytical calculations, particularly by Rees
\cite{1986MNRAS.218P..25R}, suggesting that absorption by gas confined
in dark matter ``mini-halos'' could account for the basic properties
of the \Lya~forest, two major developments in the 1990Õs allowed
astrophysicists to tackle the question with unprecedented precision:
the advent of high-resolution spectrographs and numerical cosmological
simulations. The new HIgh Resolution Echelle Spectrometer (HIRES) spectrograph
installed at the Keck Observatories on
Mauna Kea, Hawaii, provided quasar spectra with fully resolved
\Lya~forests. 
At about the same time, the Cosmic Background Explorer (COBE) satellite had recently detected small anisotropies, subsequently mapped in sharp detail by the Wilkinson Microwave Anisotropy Probe\cite{2003ApJS..148..175S}, imprinted on the cosmic microwave background when the universe was only 400,000 years old\cite{1992ApJ...396L...1S}. 
The COBE observations fueled efforts to model the growth of structure in the universe.
Did the minuscule fluctuations in the very early universe observed by COBE actually evolve into the rich structure observed by astronomers? 
Following advances in computer algorithms and technology, first-principles simulations traced the hydrodynamic evolution of cosmic structures from initial conditions motivated by the cosmic microwave background down to the current epoch [see discussion and references in \cite{1998ARA&A..36..267R}]. Zel'dovich did anticipate the cosmic network of filaments, knots, and sheets seen in the simulations. However, the detailed simulations turned the model into a quantitative one, with predictions worthy of comparison against the exacting data now available. Assuming that the \Lya~forest originated from absorption by intervening neutral hydrogen, it was then straightforward to produce mock spectra, which could directly be tested against the actual observations (Fig. \ref{lya forest}). 

The results of these analyses showed that the simulations were exactly right.  In
particular, the calculations of \cite{1996ApJ...457L..51H, 1996ApJ...457L..57K} demonstrated
correspondence between observations and theory over nearly eight
orders of magnitude in the strength of the absorbing structures, at
the time an unprecedented achievement in numerical cosmology.  The
level of agreement has only improved over the past decade as
simulations and observations have been refined, and as our
knowledge of the underlying cosmological model has become more
firm. The conclusion appears inescapable: the primordial fluctuations
seen in the cosmic microwave background did grow by five orders of magnitudes over billions of years,
\emph{precisely as calculated}, to form the observed cosmic web.

The discovery of the cosmic web in the \Lya~forest has had a broad and
profound impact beyond confirming the gravitational instability
paradigm for structure formation and explaining the basic properties
of the intergalactic medium in a cosmological context.
In current hierarchical models of structure formation, condensed
objects such as galaxies are thought to form in the denser clumps of
the cosmic web. The same numerical simulations that were so successful
in explaining quasar absorption systems can thus be used to
theoretically calculate the spatial distribution of galaxies, another
prediction that can be put to stringent test by observations, and
indeed has been spectacularly vindicated by galaxy surveys.
The intergalactic medium probed by current observations is almost
fully ionised, owing to the ultra-violet radiation field from
star-forming galaxies and the supermassive black holes powering
quasars. The imprint of the cosmic web in the \Lya~forest, arising
from the small residual fraction of neutral gas, thus also traces the
cosmic radiation content, providing a unique record of the cosmic
history of star formation and black hole growth \cite{1997ApJ...489....7R}.

Observations of the cosmic web have reached beyond astrophysics and
into fundamental physics as well. In the paradigm of structure
formation that the \Lya~forest has been
instrumental in establishing, the cosmic web is
nothing but an evolved snapshot of the tiny primordial fluctuations
produced at the very beginning, in an epoch of ``inflation'' during
which the universe expanded by more than twenty-five orders of
magnitude in size. By measuring the detailed properties of
the cosmic web, we can constrain the properties of
inflation itself and the physical mechanisms driving it
\cite{1998ApJ...495...44C, 2005ApJ...635..761M}. In addition, an
overwhelming number of lines of evidence, including galactic rotation
curves, the velocity dispersions of galaxy clusters and gravitational
lensing, indicate that the cosmic matter budget is dominated by
invisible ``dark matter.'' 

Owing to its lack of non-gravitational interactions with the rest of the universe, however, the nature of dark matter
remains largely a mystery. The leading dark matter candidate is an
exotic particle, never yet observed in Earth-bound laboratories,
generically referred to as a ``weakly interacting massive particle,''
or WIMP. Observations of large-scale structure through galaxy
surveys first ruled out the neutrino as a candidate for ``hot'' dark
matter \cite{1983ApJ...274L...1W}, leading to the present cold dark
matter paradigm, and are now putting severe upper limits on the neutrino's mass
\cite{2006JCAP...10..014S}. The \Lya~forest, being more sensitive to
smaller scales of order one megaparsec, currently provides
the most rigorous constraints on intermediate models of ``warm,'' more
massive, dark matter \cite{2006PhRvL..97g1301V, 2006PhRvL..97s1303S}.

The bulk of research on cosmic structure, in particular from the
\Lya~forest and galaxy surveys, has thus far focused on observations
in the optical part of the electromagnetic spectrum. Astronomy,
however, is an increasingly multi-wavelength endeavour, and the study
of the cosmic web is poised to follow this trend as new frontiers are
explored. Accordingly, new opportunities for novel theoretical
calculations and observational discoveries abound.

One of the most exciting frontiers of contemporary cosmology is
the early universe, when the first stars and galaxies formed
and illuminated their surroundings 
During this epoch of ``reionisation,'' 
electrons were unbound from hydrogen atoms (which had combined to become neutral at the time the anisotropies were imprinted in the cosmic microwave background) by ultra-violet radiation, which had begun filling intergalactic space.
Neutral hydrogen prior to and during reionisation can potentially
be observed through its emission of 21-cm radio radiation.
Several
low-frequency observatories -- such as the Murchison Widefield Array
(MWA) in Western
Australia, the Low Frequency Array
(LOFAR) in the Netherlands and,
ultimately, the Square Kilometer Array
(SKA) - are being planned,
constructed, or entering service to detect this redshifted
emission. Simultaneously, ever-more powerful infrared telescopes are
attempting to discover the first sources of light directly.  
This is in fact a primary scientific goal of the James Webb Space
Telescope, the Hubble Space 
Telescope's successor scheduled for launch in 2013, as well as of very large
ground-based observatories.

New windows are also now opening on the lower-redshift universe. A new
ultra-violet spectrograph, the Cosmic Origins Spectrographs
(COS), will be installed
aboard Hubble during its 2008 servicing mission, and will provide a
direct and detailed probe of the cosmic distribution of helium at
intermediate redshifts. Since more energetic photons are required
to doubly ionise helium, it is
thought to have been fully ionised later than hydrogen, near the peak
of quasar activity.
The study of helium reionisation thus promises
to become a powerful probe of quasar activity and its feedback on the
intergalactic medium in the very near future.

At still lower redshifts, the non-linear growth of cosmic structure
shocks the intergalactic medium, heating it to temperatures up to
$10^{7}$ K \cite{1999ApJ...514....1C, 2001ApJ...552..473D}. The high
temperatures in the local ``warm-hot intergalactic medium,'' or WHIM,
imply that yet heavier elements are ionised and hence higher-energy wavelengths, including X-rays, are the probes of choice for this physical regime.

Each of these observations is, however, accompanied
by theoretical challenges. The epochs of hydrogen and helium reionisation 
involve non-equilibrium radiative transfer
phenomena, which are only beginning to be included in cosmological
simulations. Simulations must evolve large regions of the universe to overcome
cosmic variance and capture the large scales of reionisation,
yet high resolution is needed to resolve
the sources and sinks of radiation, as well as the clumpiness of the
gas. At the present time, approximations are used to treat the problem
with available computational resources and explore the important
effects without resorting to prohibitive, fully self-consistent
radiation hydrodynamics \cite{2007MNRAS.377.1043M,
2007ApJ...654...12Z}. Feedback processes such as galactic winds and
metal enrichment are only crudely, if at all, included and
may be particularly important to understand the low-redshift
high-energy absorption. Moreover, star and galaxy formation and quasar
activity are often modeled using prescriptions, which
albeit physically motivated, do not offer the satisfaction of \emph{ab
initio} calculations.

As the rich array of new and diverse observations promise a wealth of
surprises, will the theoreticians be clever enough to provide results
with true predictive power for the multi-wavelength cosmic web?

\bibliography{scibib}

\bibliographystyle{Science}

\end{document}